\documentclass[conference, letterpaper]{./IEEEtran}

\ifCLASSINFOpdf
  \usepackage[pdftex]{graphicx}
\else
 
\fi
\usepackage{amsmath}
\usepackage{amssymb}
\usepackage{amsfonts}
\usepackage{subfig}
\usepackage{multirow}
\usepackage{multicol}
\usepackage{subfig}
\usepackage{cases}
\usepackage{color}

\let\oldsqrt\sqrt
\def\sqrt{\mathpalette\DHLhksqrt}
\def\DHLhksqrt#1#2{%
\setbox0=\hbox{$#1\oldsqrt{#2\,}$}\dimen0=\ht0
\advance\dimen0-0.2\ht0
\setbox2=\hbox{\vrule height\ht0 depth -\dimen0}%
{\box0\lower0.4pt\box2}}

\begin{document}

\title{Approaching the Gaussian channel capacity with APSK constellations}

\author{
\IEEEauthorblockN{Hugo M{\'e}ric}
\IEEEauthorblockA{NIC Chile Research Labs - Santiago, Chile\\
Email: hmeric@niclabs.cl}
}

\maketitle

\begin{abstract}
We consider the Gaussian channel with power constraint $P$. A gap exists between the channel capacity and the highest achievable rate of equiprobable uniformly spaced signal. Several approaches enable to overcome this limitation such as constellations with non-uniform probability or constellation shaping. In this letter, we focus on constellation shaping. We give a construction of amplitude and phase-shift keying (APSK) constellations with equiprobable signaling that achieve the Gaussian capacity as the number of constellation points goes to infinity.  
\end{abstract}

\IEEEpeerreviewmaketitle

\section{Introduction}

We consider the additive white Gaussian noise (AWGN) channel with signal-to-noise ratio (SNR) $P/N_0$, where $P$ is the average power constraint of the input signal and $N_0$ the noise variance. Let $\mathrm{snr}=P/N_0$, the Gaussian channel capacity is
\begin{equation}
C = \frac{1}{2} \log_2\left(1+\mathrm{snr}\right) \text{ bits/dimension},
\end{equation}
and the optimal input distribution is Gaussian with zero mean and variance $P$ \cite{cover91}. If the channel inputs are also subject to peak power constraints, the capacity and the optimal input distribution were studied for scalar and quadrature channels in \cite{Smith71} and \cite{shamai95}, respectively. In two dimensions, the optimal distribution is discrete along the radial direction and continuous in the angular direction. This proves
the advantages of circular APSK constellations under those power constraints. Even if we do not consider peak power constraints in our work, this shows the importance of designing APSK modulations.

In most practical systems, the input signal is \emph{uniformly distributed} on a finite set of points called the constellation. For the Gaussian channel, the constellation design generally focuses on maximising the squared Euclidean distance between the signal points under the power constraint. This results in (uniformly spaced) rectangular constellations. At large SNR, these signal sets exhibit a gap of $\frac{\pi \mathrm{e}}{6} \approx 1.56$ dB with the Gaussian capacity \cite{forney84}. Hence rectangular constellations with equiprobable signaling cannot achieve the Gaussian capacity.

Several approaches enable to overcome this limitation such as constellation shaping, constellations with non-uniform probabilities \cite{kschischang93} or lattice encoding and decoding \cite{erez04}. In this paper, we focus on the former solution. Sun and van Tilborg were the first to  present a sequence of random variables equiprobably distributed over a finite support that achieves the Gaussian capacity (in one dimension) as the signal set cardinality tends to infinity \cite{sun93}. This work was extended by Schwarte that provided sufficient conditions for uniform input distributions with finite support to approach the Gaussian capacity in any dimension \cite{schwarte96}. Similar conditions are given in \cite[Theorem 9]{wu10-2}.

A recent work by Wu and Verd\'u studied the AWGN channel capacity with finite signal set \cite{wu10}. They showed that as the input signal cardinality grows, the constellation capacity approaches the Gaussian capacity exponentially fast. They also introduced a family of constellation, based on the Hermite polynomials roots, achieving exponential convergence. The resulting modulations combine constellation shaping with non-uniform probabilities.

In this work, our main contribution (presented in Section~\ref{part2}) is the construction of APSK constellations with equiprobable signaling that achieve the Gaussian capacity as the number of constellation points goes to infinity. The constellation design relies on the Box-Muller theorem \cite{box58}. Designing APSK constellations is an important topic in communications. For instance, De Gaudenzi \textit{et al.} addressed the optimisation of APSK constellations for nonlinear channel in \cite{gaudenzi06}. Indeed, this may benefit some practical systems as the digital video broadcasting (DVB) standards that implement APSK constellations with large cardinality, up to 256 \cite{s2x}.

\section{APSK signals approach the Gaussian capacity}\label{part2}

The Box-Muller transform is a method for generating pairs of independent normally distributed random numbers \cite{box58}. The method proposed by Box and Muller works as follows: let $U$ and $V$ be two independent random variables that are uniformly distributed in the interval $(0, 1)$. Consider the random variables
\begin{equation}
X = \sqrt{-2 \log_\mathrm{e} U} \cos \left(2 \pi V\right)
\end{equation}
and
\begin{equation}
Y = \sqrt{-2 \log_\mathrm{e} U} \sin \left(2 \pi V\right),
\end{equation}
then we have the following result \cite{box58}:\\
\textbf{Theorem (Box-Muller).} \textit{$X$ and $Y$ are independent random variables with a standard normal distribution.}

Based on the Box-Muller transform, we show how to design APSK constellations that achieve the Gaussian capacity as the number of constellation points goes to infinity. Let $n \geqslant 1$ be an integer, consider $U_n$ and $V_n$ two random variables uniformly distributed on 
\begin{equation}
\left\{\, \frac{1}{2n} + \frac{k}{n} \bigm| 0 \leqslant k \leqslant n-1 \,\right\}.
\label{set}
\end{equation}
Then we define $\varphi \colon (0,1)^2 \to \mathbb{R}^2$ by
\begin{equation*}
\varphi (x , y) = \left( \sqrt{-P\log_\mathrm{e}x} \cos(2 \pi y) , \sqrt{-P\log_\mathrm{e}x} \sin(2 \pi y) \right).
\end{equation*}
Finally, we introduce the random vector $W_n = \varphi \left( U_n, V_n \right)$. By construction, $W_n$ is a random vector \emph{uniformly distributed} on a set $\mathcal{C}_n$ of $n^2$ points in $\mathbb{R}^2$. The points in $\mathcal{C}_n$ are distributed on $n$ circles, each circle containing $n$ points. The set (\ref{set}) ensures that $U_n$ is never equal to zero, avoiding problem with the logarithm, and also that the constellation points are uniformly distributed on each circle.

Before stating our result, we present a lemma that will be helpful to prove that $\mathcal{C}_n$ satisfies the average power constraint.\\
\textbf{Lemma.} \textit{For any integer $k \geqslant 1$,
\begin{equation}
k \log_\mathrm{e} k - k \leqslant \sum_{j=0}^{k-1} \log_\mathrm{e} \left( j + \frac{1}{2} \right).
\end{equation}}
\begin{IEEEproof}
Let $t \geqslant 1/2$ be a real number, it is easy to verify that $\int_{t-\frac{1}{2}}^{t+\frac{1}{2}} \log_\mathrm{e} u \mathrm du \leqslant \log_\mathrm{e} t$ . This leads to
\begin{align}
 k \log_\mathrm{e} k - k & = \int_{0}^{k} \log_\mathrm{e} u \mathrm du 						\nonumber\\
			 & = \sum_{j=0}^{k-1} \int_{j}^{j+1} \log_\mathrm{e} u \mathrm du 			\nonumber\\
			 & \leqslant \sum_{j=0}^{k-1} \log_\mathrm{e} \left( j + \frac{1}{2} \right).
\end{align}
\end{IEEEproof}

\begin{figure*}[!t]
\centerline{\subfloat[$\mathrm{snr}=5$ dB]{\includegraphics[width=0.32\textwidth]{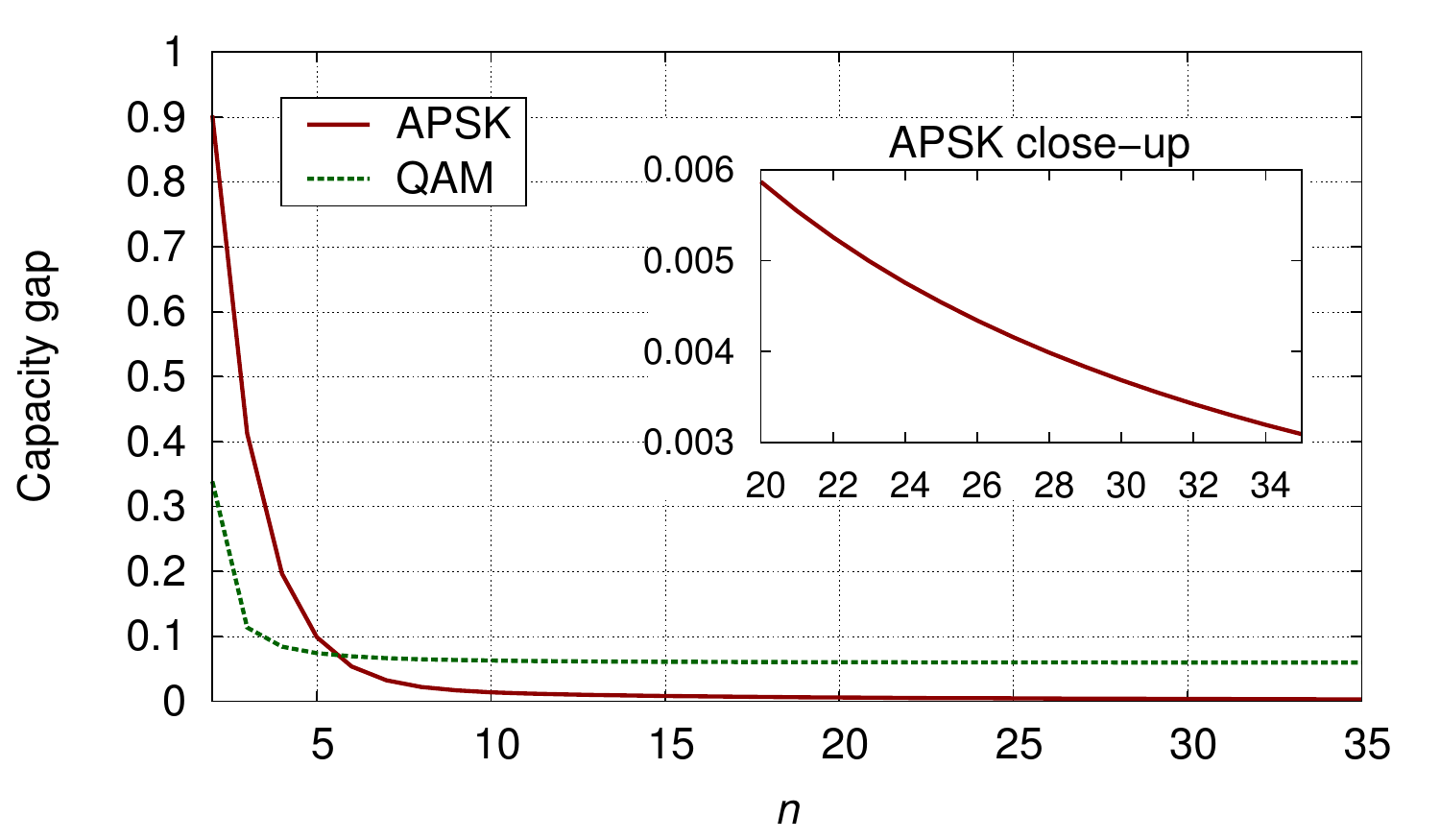}%
\label{capacity_gap_5db}}%
\hfil
\subfloat[$\mathrm{snr}=10$ dB]{\includegraphics[width=0.32\textwidth]{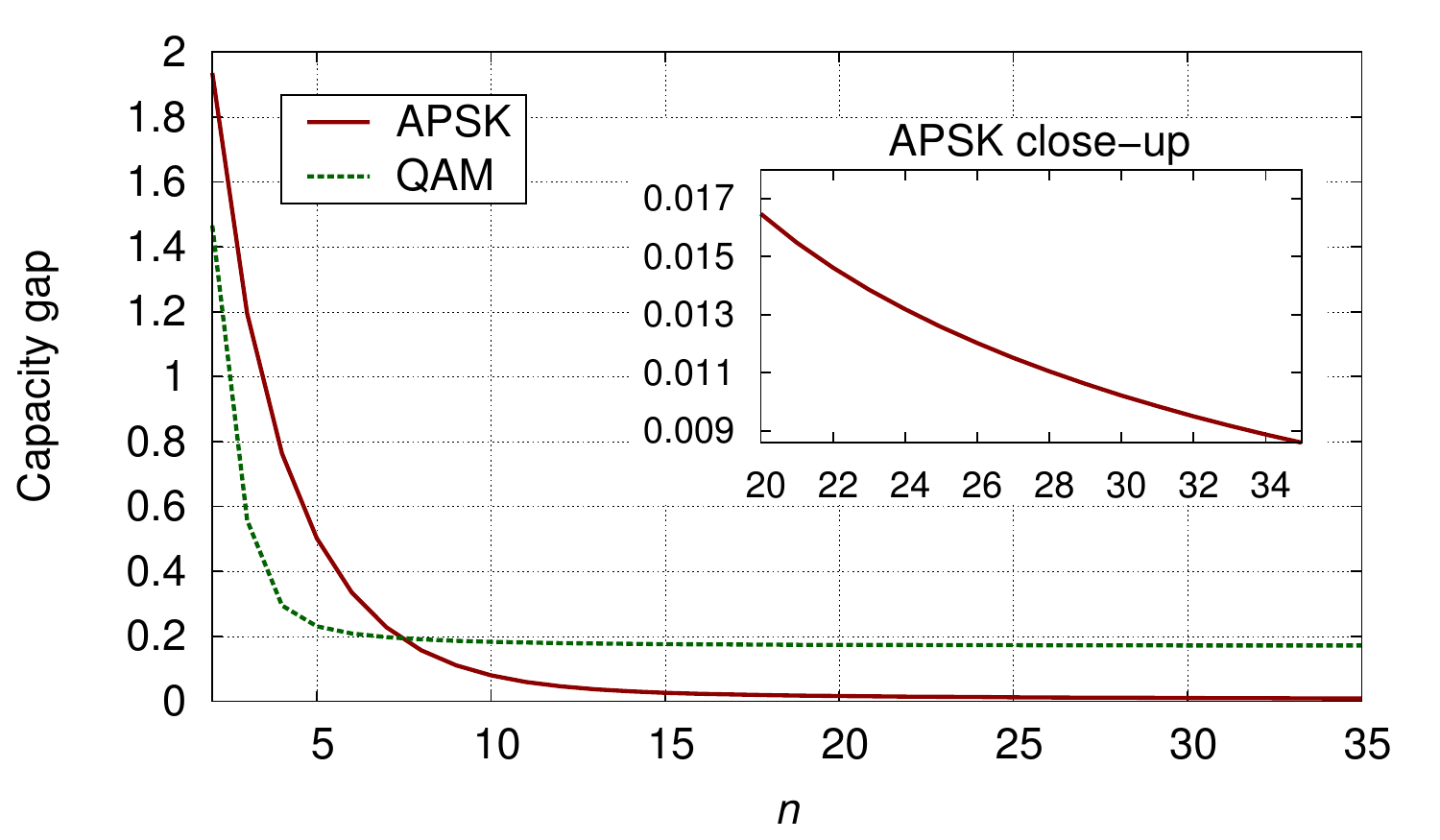}%
\label{capacity_gap_10db}}%
\hfil
\subfloat[$\mathrm{snr}=15$ dB]{\includegraphics[width=0.32\textwidth]{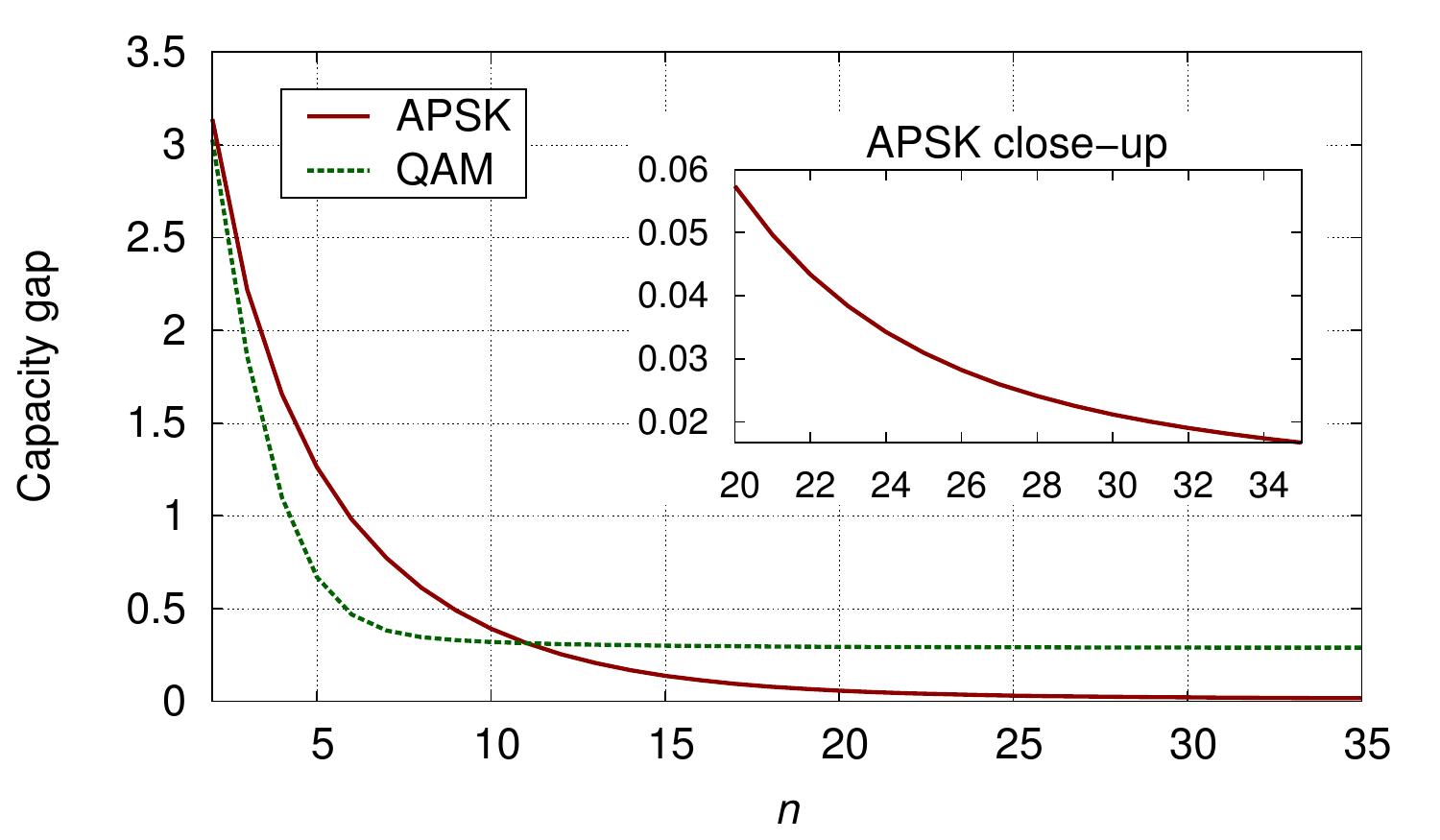}%
\label{capacity_gap_15db}}}%
\caption{Gap to Gaussian channel capacity for the regular QAM and the proposed APSK ($2 \leqslant n \leqslant 35$)}
\label{capacity_gap}
\end{figure*}

The following theorem is the main result of the paper:\\
\textbf{Theorem.} \textit{The APSK constellations previously designed achieve the Gaussian channel capacity as $n \rightarrow \infty$. More formally,
\begin{equation}
I(W_n; W_n+N) \underset{n\to\infty}{\longrightarrow} \log_2(1+\mathrm{snr}),
\end{equation}
where $N \sim \mathcal{N}(0, N_0)$ is the AWGN with variance $N_0$ ($N_0/2$ on each dimension).}  
\begin{IEEEproof}
The idea is to show that the sequence $(W_n)_{n\geqslant1}$ satisfies the two conditions given in \cite[Corollary~7]{schwarte96}.
 
\underline{Condition 1:} First, we prove that the constellation $\mathcal{C}_n$ satisfies the power constraint for $n \geqslant 1$. The input signal energy is
\begin{align}
\mathbb{E}\left[W_n^2\right] & = \frac{1}{n^2} \sum_{\mathbf{w} \in \mathcal{C}_n } \| \mathbf{w} \|^2 							\nonumber\\
                	     & = \frac{1}{n} \sum_{k=0}^{n-1} -P \log_\mathrm{e} \left( \frac{1}{2n} + \frac{k}{n} \right)	  			\nonumber\\
			     & = -\frac{P}{n} \left( -n \log_\mathrm{e} n +  \sum_{k=0}^{n-1} \log_\mathrm{e} \left( k + \frac{1}{2} \right) \right)	\nonumber\\
			     & \leqslant P, 														\label{nrj}
\end{align}
where the last inequality follows from the previous lemma. Thus the constellation $\mathcal{C}_n$ verifies the power constraint.

\underline{Condition 2:} Next, we show that the distribution of $W_n = \varphi(U_n, V_n)$ converges to the distribution of a Gaussian variable, denoted $W^*$, with variance $P/2$ on each dimension. To that end, we study $\Phi_{W_n}$ the characteristic function of $W_n$. By definition,
\begin{equation}
\Phi_{W_n}(\mathbf{t}) = \mathbb{E} \left[ \mathrm{e}^{i \langle \mathbf{t} | W_n \rangle} \right],
\label{characteristic}
\end{equation}
where $\mathbf{t}=(t_1,t_2) \in \mathbb{R}^2$ and $\langle \cdot | \cdot \rangle$ is the scalar product. Introducing the function
\begin{equation}
\psi (x,y) = \mathrm{e}^{i \sqrt{-P \log_\mathrm{e} x} \left( t_1  \cos ( 2\pi y) + t_2  \sin (2 \pi y) \right)},
\end{equation}
we can write (\ref{characteristic}) as
\begin{equation*}
\Phi_{W_n}(\mathbf{t}) = \frac{1}{n^2} \sum_{k=0}^{n-1} \sum_{l=0}^{n-1} \psi \left( \frac{1}{2n} + \frac{k}{n}, \frac{1}{2n} + \frac{l}{n} \right).
\end{equation*}
Moreover, it results from the Box-Muller theorem that
\begin{equation}
\Phi_{W^*}(\mathbf{t}) = \iint_{(0,1)^2} \psi (x,y) \mathrm dx \mathrm dy.
\label{characteristic_Wopt}
\end{equation}

Now, we consider the sequence $(\psi_n)_{n \geqslant 1}$ defined by
\begin{equation}
\psi_n (x,y) = \psi \left( \frac{1}{2n}+\frac{\lfloor nx \rfloor}{n} , \frac{1}{2n}+\frac{\lfloor ny \rfloor}{n} \right),
\end{equation}
where $\lfloor \cdot \rfloor$ is the floor function. The sequence $(\psi_n)_{n \geqslant 1}$ converges pointwise to $\psi$. Over $(0,1)^2$, $|\psi_n|$ is dominated by the constant function equals to 1 (for all $n$). Applying the Lebesgue's dominated convergence theorem \cite[Theorem 16.4]{billingsley}, we obtain
\begin{equation}
\iint_{(0,1)^2} \psi_n (x,y) \mathrm dx \mathrm dy \underset{n\to\infty}{\longrightarrow} \iint_{(0,1)^2} \psi (x,y) \mathrm dx \mathrm dy.
\label{lebesgue}
\end{equation}
Moreover,
\begin{align}
\iint_{(0,1)^2} \psi_n (x,y) \mathrm dx \mathrm dy & = \sum_{k=0}^{n-1} \sum_{l=0}^{n-1} \int_{\frac{k}{n}}^{\frac{k+1}{n}} \int_{\frac{l}{n}}^{\frac{l+1}{n}} \psi_n (x,y) \mathrm dx \mathrm dy \nonumber\\
						   & = \frac{1}{n^2} \sum_{k=0}^{n-1} \sum_{l=0}^{n-1} \psi \left( \frac{1}{2n} + \frac{k}{n}, \frac{1}{2n} + \frac{l}{n} \right)      \nonumber\\
						   & = \Phi_{W_n}(\mathbf{t}). 
\label{int_sum}
\end{align}
Combining (\ref{characteristic_Wopt}), (\ref{lebesgue}) and (\ref{int_sum}), we obtain
\begin{equation}
\Phi_{W_n}(\mathbf{t}) \underset{n\to\infty}{\longrightarrow} \Phi_{W^*}(\mathbf{t}).
\end{equation}
Finally, it follows from the continuity theorem that $W_n$ converges weakly to $W^*$ \cite[Theorem 26.3]{billingsley}. 

\underline{Conclusion:} The constellation uniformly distributed over $\mathcal{C}_n$ satisfies both conditions in \cite[Corollary~7]{schwarte96}. We conclude that the constellations approach the Gaussian channel capacity with power constraint $P$ as $n \rightarrow \infty$. 
\end{IEEEproof}

To illustrate the previous theorem, \figurename~\ref{capacity_gap} depicts the gap to Gaussian capacity for the regular QAM and the proposed APSK constellations for various SNR values. As expected, the capacity gap vanishes for the APSK as $n \rightarrow \infty$. However the convergence speed remains an open question.

\section{Concluding remarks}\label{part3}

Our APSK design enables to achieve the Gaussian capacity as the constellation size grows to infinity. In that sense, APSK modulations are better than  regular QAM. However, non-uniform QAM with equiprobable signaling may also reach the Gaussian capacity. Indeed, a way to construct such modulations is to consider a capacity-achieving signal in one dimension (for instance \cite{sun93} or \cite{schwarte96}) and transmit such signals on quadrature carriers.

\begin{figure}[!ht]
\centering
\includegraphics[width = 0.7\columnwidth]{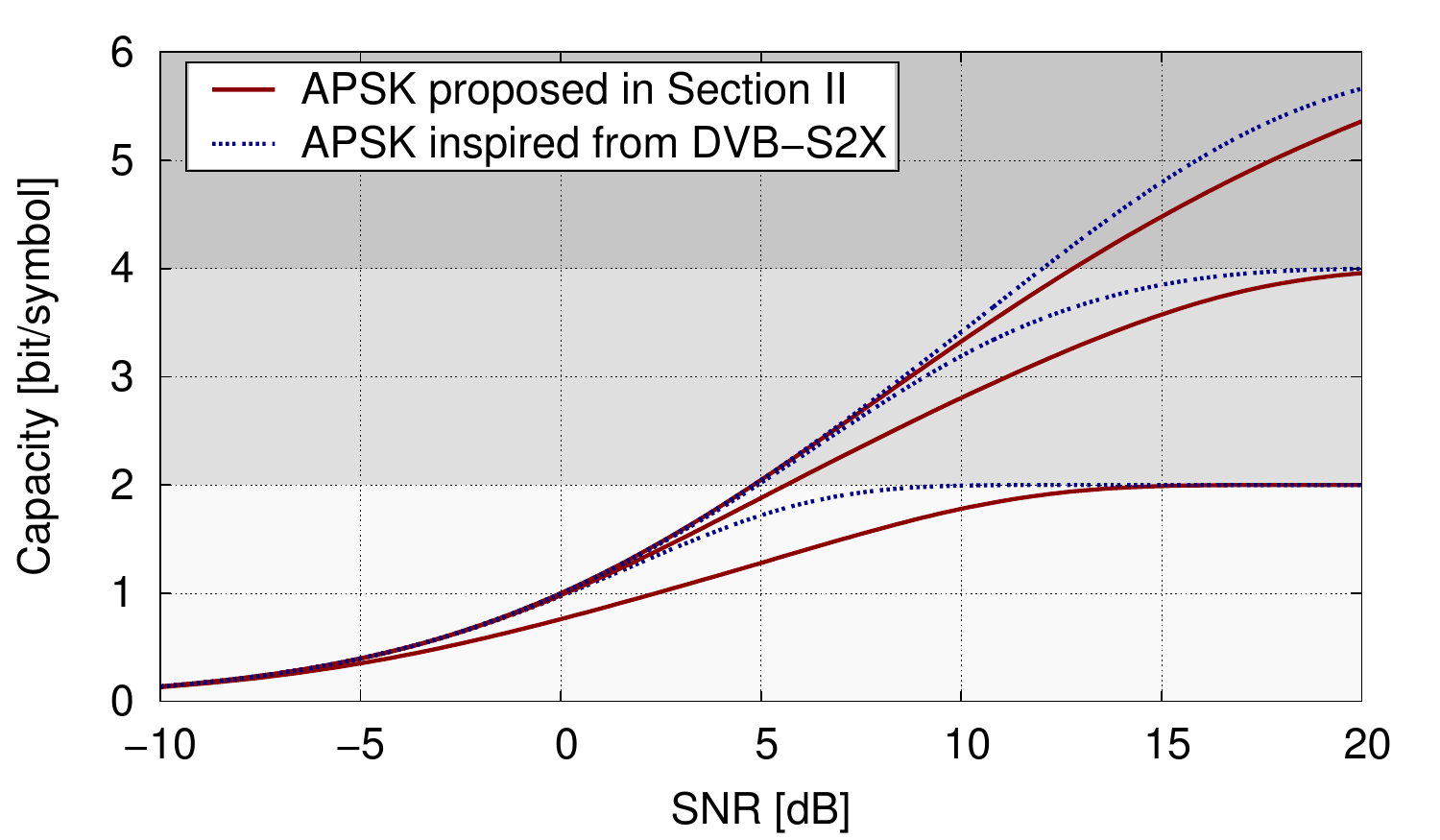}
\caption{Achievable rate comparison of two capacity-achieving APSK for $n=2, 4$ and 8 (i.e., 4, 16 and 64 APSK)}
\label{comparison}
\end{figure}

The DVB-S2X standard implements various APSK modulations, but no justification about the constellation design is provided \cite{s2x}. When the constellation cardinality is of the form $2^{2k}$ (i.e., $n=2^k$ with the previous notations), the standard splits the points on $n/2$ circles where each circle contains $2n$ points uniformly distributed. Including these conditions within our framework, this is equivalent to consider $U_n$ and $V_n$ uniformly distributed on
\begin{equation}
\left\{\, \frac{1}{n} + \frac{2k}{n} \bigm| 0 \leqslant k \leqslant \frac{n}{2}-1 \,\right\}
\end{equation}
and
\begin{equation}
\left\{\, \frac{1}{4n} + \frac{k}{2n} \bigm| 0 \leqslant k \leqslant 2n-1 \,\right\},
\end{equation}
respectively. The proof in Section~\ref{part2} may be adapted to show that such constellations also reach the Gaussian capacity as $n \rightarrow \infty$. \figurename~\ref{comparison} compares both constellations in terms of achievable rates for various $n$. Even if they asymptotically achieve the Gaussian capacity, the second design performs better. Moreover this scheme reduces the peak-to-average power ratio, a suitable property for practical systems.

As we can see, several approaches are possible to construct capacity-achieving APSK. We also believe that rotating in the complex plane a one dimensional capacity-achieving constellation should work. In practice, other criteria such as the peak-to-average power ratio should be considered to select the best constellation. 

For scalar AWGN channel with input cardinality $m$, Wu and Verd\'u showed that the achievable rate approaches exponentially fast the Gaussian capacity as $m$ grows \cite{wu10}. The convergence speed of the proposed APSK is an open question.

\ifCLASSOPTIONcaptionsoff
  \newpage
\fi

\nocite{*}
\bibliographystyle{IEEEtran}
\bibliography{biblio}

\end{document}